\documentclass[runningheads]{llncs}
\usepackage{physics}
\usepackage{algorithm}
\usepackage[noend]{algpseudocode}
\usepackage{graphicx}
\usepackage{subfig}

\begin{document}
\title{Investigating Bell Inequalities for Multidimensional Relevance Judgments in Information Retrieval}
%
%
\author{Sagar Uprety\inst{1}\orcidID{0000-0001-7858-6265} \and
Dimitris Gkoumas\inst{1} \orcidID{0000-0002-0524-0332} \and
Dawei Song\inst{1,2} \orcidID{0000-0002-8660-3608}}

\authorrunning{S. Uprety et al.}
%
\institute{The Open University, Milton Keynes, UK \\
\email{\{sagar.uprety, dimitris.gkoumas, dawei.song\}@open.ac.uk}\\
\and
Beijing Institute of Technology, Beijing, China
}
\maketitle              
\begin{abstract}
Relevance judgment in Information Retrieval is influenced by multiple factors. These include not only the topicality of the documents but also other user oriented factors like trust, user interest, etc. Recent works have identified and classified these various factors into seven dimensions of relevance. In a previous work, these relevance dimensions were quantified and user's cognitive state with respect to a document was represented as a state vector in a Hilbert Space, with each relevance dimension representing a basis. It was observed that relevance dimensions are incompatible in some documents, when making a judgment. Incompatibility being a fundamental feature of Quantum Theory, this motivated us to test the Quantum nature of relevance judgments using Bell type inequalities. However, none of the Bell-type inequalities tested have shown any violation. We discuss our methodology to construct incompatible basis for documents from real world query log data, the experiments to test Bell inequalities on this dataset and possible reasons for the lack of violation.

\end{abstract}

\section{Introduction}
Information Retrieval (IR) is defined as finding material (documents, videos, audio, etc.) of an unstructured nature that are relevant to an information need of the user. Information Need (IN) of a user is usually expressed as a query. An essential component of IR is the concept of relevance of documents. It is defined as how well a document satisfies the user Information Need. Relevance in IR was traditionally considered to be Topical, i.e. how well the content of the retrieved document matches the topic of the query(e.g. text match). As content similarity matching techniques have become more accurate, almost all of the documents obtained for a query generally satisfy the topicality criteria. Hence users tend to consider other factors while judging documents. These different factors have been investigated in several works~\cite{ASI:Barry-relevance,ASI:Xu-MURM,MURM-psychometrics}. In \cite{ASI:Jingfei}, seven relevance dimensions were identified. Each of these dimensions was quantified by defining certain features, which could be extracted from all query-document pairs. These seven dimensions are described in Table 1. 

\begin{table}[t!]
      \caption{Seven dimensions of relevance}
\label{table-rel-dim}
\begin{center}
    \begin{tabular}{  l  p{9cm} }
    \hline
Relevance Dimensions 
\\ \hline
Topicality & The extent to which the retrieved document is related to the topic of the current query.
\\ 
Reliability &  The degree to which the content of the document is true, accurate and believable. Determined by the reliability of source.
\\ 
Understandability &  Extent to which the contents are readable. Vocabulary, complexity of sentences, layout of pages, etc. taken into consideration.
\\
Interest & Topics from user’s past searches.
\\
Habit & Focus on behavioral preference of users, e.g. always using certain websites for particular tasks.
\\
Scope & Whether both breadth and depth of the document are suitable to the Information Need	
\\
Novelty & Whether the document contains information which is new to the user, or the document itself is newly created
\\
\hline
    \end{tabular}
\end{center}
\end{table}

In \cite{sigir}, a document defined using these relevance dimensions is represented as a two dimensional Hilbert space. Each of the seven relevance dimensions is represented as a basis. The different basis correspond to the different perspectives of relevance judgment for the same document. Based on which relevance dimension is considered, the same document will have different probabilities of relevance. Thus the document exists in multiple states (e.g. highly relevant, not relevant, moderately relevant, etc.) simultaneously and we get a particular judgment depending upon which criteria (relevance dimension) the user used to judge (measure) it. This is analogous to the measurement of electron spin which is either up or down in direction, but depends upon which axis it is measured in. Electrons with spin up along the Z-axis may have both up and down components along the X-axis. So a document may look relevant based on the Topicality dimension, but may not be so along, say, the Reliability dimension. We discuss the methodology used to quantify these seven dimensions and construct Hilbert spaces for documents in the next section.

This incompatibility in judgment perspectives is a fundamental feature of Quantum Mechanics~\cite{sakurai}. Incompatibility forbids the possibility of jointly determining the outcome of an event from two perspectives. We investigate whether decision making in IR, consisting of multiple perspectives, has an analogous quantum phenomena. A formal test of quantumness of systems was given in 1964 by John Bell~\cite{bell}. He formulated an inequality which cannot be violated by classical systems governed by joint probability distributions. Quantum Mechanics was shown to violate it for particular settings. In this work, we use another version of the Bell inequality, called the CHSH inequality~\cite{chsh}. The CHSH inequality is given by equation (\ref{CHSH}) for two systems $A$ and $B$ where observables $A_1$ and $A_2$ can be measured in system $A$ and $B_1$ and $B_2$ can be measured in system $B$. $A_i$ and $B_i$ can take values only in $\{\pm1\}$.  It is assumed that the observables have pre-existing values which are not influenced by any other measurement. 

\begin{equation}
|\langle A_1B_1 \rangle +\langle A_1B_2 \rangle+\langle A_2B_1 \rangle-\langle A_2B_2 \rangle| \le 2 \label{CHSH}
\end{equation}
The CHSH inequality is violated in Quantum Mechanics using a special composite state of two systems, called the Bell state~\cite{Nielsen}, which has the following form:
\begin{equation}
\ket{\psi} = \frac{1}{\sqrt[]{2}}(\ket{00} + \ket{11})
\end{equation}
where $\ket{0}$ and $\ket{1}$ represent the standard basis for the two systems. Initially, both the systems are in a superposed state. The two outcomes, i.e., corresponding to the $\ket{0}$ and $\ket{1}$ vectors can be obtained with equal probabilities. However, on measuring one system, if one obtains the outcome corresponding to the basis vector $\ket{0}$, the state of the composite system collapses to $\ket{00}$. Now it is known for certain that the outcome of the second system also corresponds to $\ket{0}$. This is true even if the two systems are spatially separated - the measurement on one system reveals the state of the other, instantaneously. 

Violation of Bell inequalities by such entangled states prove the impossibility of the existence of a joint probability distribution for the two systems. It rules out the concept of "Local Realism" of the classical world, which is the assumption made while deriving the Bell inequalities. 'Local' implies the fact that measurement of one system does not influence that of a spatially separated system. 'Realism' assumes that values of physical properties of systems have definite values and exist independent of observation~\cite{Nielsen}.

There have been several works which have investigated violation of Bell inequalities in macroscopic and cognitive systems~\cite{Aerts-macrosco-2000,Aerts-sozzo-entang-2013,Bruza-entangled-2011}. This work also investigates the Bell inequalities for violation by user's composite state for judgment of two documents. After describing the methodology used to quantify the seven relevance dimensions, we describe equivalent Bell inequalities for the user states for documents. Subsequently we give details of the experimental settings used to form the composite system of documents.

\section{Quantifying Relevance Dimensions}

We represent each document as a two-dimensional real valued Hilbert space. The two basis vectors correspond to relevance and non-relevance of a dimension. For the seven dimensions, we have seven different basis in the Hilbert space. The user's cognitive state for this document is a vector in the Hilbert space, a superposition of the basis vectors. Using the Dirac notation, we get the user state for a document d in different basis as:

\begin{align}
\ket{d} &= \alpha_{11}\ket{R_{hab}} + \beta_{11}\ket{\widetilde{R_{hab}}} \nonumber \\
	      &= \alpha_{12}\ket{R_{int}} + \beta_{12}\ket{\widetilde{R_{int}}}
\end{align}
and so on, in all seven basis. The coefficients $|\alpha_x|^2$ is the weight (i.e., probability of relevance) the user assigns to document $d$ in terms of the dimension $x$ , and $|\alpha_x|^2 + |\beta_x|^2 = 1$. 

To calculate the coefficients of superposition in a basis, we use the same technique as \cite{sigir18}. The dataset is of query logs from the Bing search engine. Following the methodology in \cite{ASI:Jingfei}, we define a set of features for each of the seven relevance dimensions. For each query-document pair, the set of features for each dimension are extracted and integrated into the LambdaMART~\cite{Burges2010FromRT} Learning to Rank (LTR) algorithm to generate seven relevance scores (one for each dimension) for the query-document pair. Due to lack of space, we refer the readers to \cite{ASI:Jingfei} for more details on the features defined for each dimension and also how they are used in the LTR algorithm. We thus get seven different ranked lists for a query, corresponding to each relevance dimension. Then the scores assigned to a document for each dimension are normalized using the min-max normalization technique, across all the documents for the query. The normalized score for each dimension forms the coefficient of superposition of the relevance vector for the respective dimension. For example, for a query $q$, let $d_1, d_2, ..., d_n$ be the ranking order corresponding to the "Reliability" dimension, based on the relevance scores of $\lambda_1, \lambda_2, ..., \lambda_n$ respectively. We construct the vector for document $d_1$ in the 'Reliability' basis as:
\begin{equation}
\ket{d_1} = \alpha_{11}\ket{R_{rel}} +\beta_{11}\ket{\widetilde{R_{rel}}}
\end{equation} 
where $\alpha_{11} = \sqrt{\frac{\lambda_1 - min(\lambda)}{max(\lambda) - min(\lambda)}}$, where $max(\lambda)$ is the maximum value among $\lambda_1, \lambda_2, ..., \lambda_n$. Square root is taken to enable calculation of probabilities according to the Born rule. We can thus represent this document in all the seven basis and therefore all the documents in their respective Hilbert spaces.

For documents where $\alpha_{11}$ and $\alpha_{12}$ are different, we get incompatible basis. Incompatibility in relevance dimensions for judging documents can be manifested in terms of Order Effects. Different order of considering relevance dimensions while judging a document will lead to different final judgments. As an example, consider a document with the following Hilbert space \cite{Uprety-ictir}:
\begin{align}
\ket{d} &= 0.9715\ket{R} +0.2370\ket{\widetilde{R}}\\
 		  &= 0.3535\ket{T} +0.9354\ket{\widetilde{T}}
\end{align}
We take the $Reliability$ basis($\ket{R},\ket{\widetilde{R}}$)as the standard basis. Representing $Topicality$ basis in the standard $Reliability$ basis, we get (Appendix A):
\begin{equation}
\ket{T} = 0.5651\ket{R} + 0.8250\ket{\widetilde{R}}
\end{equation}

\begin{figure}
    \subfloat[]{
        \includegraphics[width=0.5\textwidth]{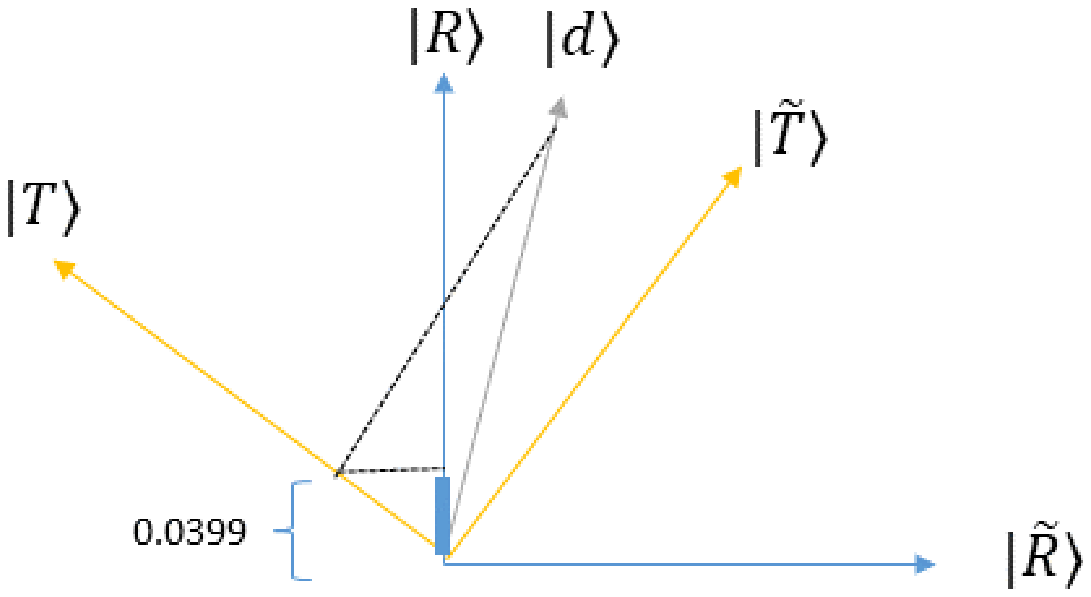}}
       \subfloat[]{
        \includegraphics[width=0.5\textwidth]{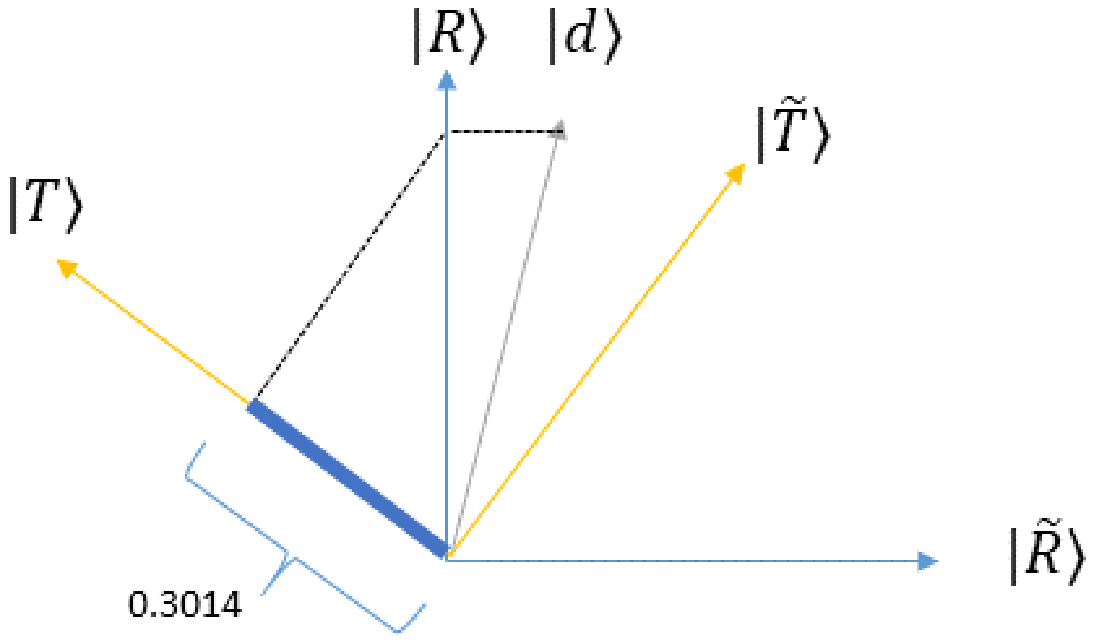}}
        \caption{Hilbert Space Representation of Order Effects}
        \label{fig-order-effects}
\end{figure}
Suppose that while judging Document $d$, the user has the order $Topicality\rightarrow Reliability$ in mind. Then the final probability of relevance is the projection from $d \rightarrow T\rightarrow R$ as shown in Figure \ref{fig-order-effects}.a. This is calculated as $|\bra{T}\ket{d}|^2|\bra{R}\ket{T}|^2 = 0.3535^2*0.5651^2 = 0.0399$. If the user reverses the order of relevance dimensions considered while judging document $d$, we get $d \rightarrow R\rightarrow T$ = $|\bra{R}\ket{d}|^2|\bra{T}\ket{R}|^2 = 0.9715^2*0.5651^2 = 0.3014$, which is $7.5$ times larger (Figure \ref{fig-order-effects}.b).

Order Effects in decision making have been successfully modeled and predicted using the Quantum framework~\cite{Trueblood2011,10.3389/fpsyg.2014.00612}.
\section{Deriving a Bell Inequality for Documents}
\subsection{CHSH Inequality} \label{chsh-normal}
\begin{figure}[h]
\centering
    \includegraphics[width=0.8\textwidth]{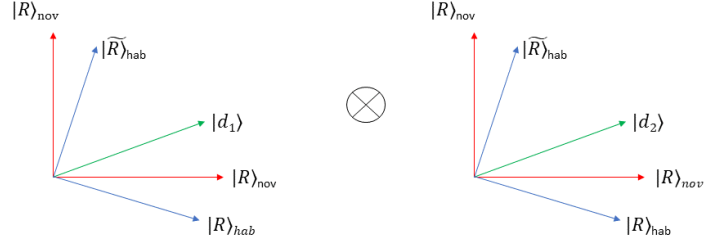}
    \caption{Tensor Product of two Hilbert Spaces}
        \label{fig-tensor-product}
\end{figure}

In section 2, we showed how we can calculate the relevance probabilities of a document for different dimensions. We constructed a Hilbert space for each document, consisting of seven different basis, representing each dimension of relevance. Two or more such documents can be considered as a composite system by taking a tensor product of the document Hilbert spaces. If $\ket{d_1}$ and $\ket{d_2}$ are the state vectors of two documents, we can represent the tensor product as $\ket{d_1} \bigotimes \ket{d_2}$. Figure \ref{fig-tensor-product} shows the geometrical representation of two such Hilbert spaces. Here $\ket{R}_{hab}$ represents Relevance in the Habit basis, or in IR terms, relevance of document $d$ with respect to the Habit dimension. Similarly, $\ket{\widetilde{R}}_{hab}$ represents irrelevance in the Habit basis. 

In the CHSH inequality, we have observables $A_1$ and $A_2$ for a system taking values in ${\pm{1}}$. For a document $d_1$, we have observables corresponding to the different relevance dimensions. Taking the case of two relevance dimensions, Habit and Novelty, we have observables $R_{hab}$ and $R_{nov}$ which take values in ${\pm{1}}$. Where $R_{hab} = +1$ corresponds to a projection on the basis vector $\ket{R}_{hab}$, $R_{hab} = -1$ corresponds to the projection on its orthogonal basis vector $\ket{\widetilde{R}}_{hab}$. 

Taking two documents as a composite system, we can write the CHSH inequality in the following way:

\begin{equation}
|\langle R_{hab1}R_{hab2}\rangle + \langle R_{hab1}R_{nov2}\rangle + \langle R_{nov1}R_{hab2}\rangle - \langle R_{nov1}R_{nov2}\rangle| \leq 2 
\end{equation}
Where the subscripts $1$ and $2$ denote that the observables belong to document $1$ and document $2$ respectively. Using the fact that $\langle AB\rangle = 1*P(AB=1) + (-1)*P(AB=-1)$ and $P(AB=1) + P(AB=-1) = 1$, we can convert the above inequality into its probability form as: 

\begin{align}
1 \hspace{0.3cm} \leq \hspace{0.3cm} &P(R_{hab1}R_{hab2} = 1) + P(R_{hab1}R_{nov2} = 1) + \\ \nonumber
& P(R_{nov1}R_{hab2} = 1) + P(R_{nov1}R_{nov2} = -1) \hspace{0.3cm} \leq \hspace{0.3cm} 3
\end{align}

We don't have the joint probabilities $P(AB)$ in our dataset, hence we assuming $P(AB) = P(A)P(B)$ (this where the assumption of realism is incorrectly made, which will not lead to the CHSH inequality violation), we get:
\begin{align}
1 \hspace{0.3cm} \leq \hspace{0.3cm} &P(R_{hab1} = 1)P(R_{hab2} = 1) + P(R_{hab1} = -1)P(R_{hab2} = -1) + \\ \nonumber
&P(R_{hab1} = 1)P(R_{nov2} = 1) + P(R_{hab1} = -1)P(R_{nov2} = -1) + \\ \nonumber
&P(R_{nov1} = 1)P(R_{hab2} = 1) + P(R_{nov1} = -1)P(R_{hab2} = -1) + \\ \nonumber
&P(R_{nov1} = 1)P(R_{nov2} = -1) + P(R_{nov1} = -1)P(R_{nov2} = 1) \hspace{0.3cm} \leq \hspace{0.3cm} 3
\end{align}
\\
As we mentioned above, $R_{hab} = +1$ corresponds to the basis vector $\ket{R_{hab}}$ and therefore $P(R_{hab1} = 1)$ corresponds to the probability that document $d_1$ is relevant with respect to the $Habit$ dimension of relevance. Therefore we can calculate these probabilities as projections in the Hilbert space:

\begin{align}
&P(R_{hab1} = 1)  \hspace{0.25cm}= |\bra{R_{hab}\ket{d_1}}^2 \\ \nonumber
&P(R_{hab1} = -1) = |\bra{\widetilde{R_{hab}}\ket{d_1}}^2 \\ \nonumber
&P(R_{nov1} = 1)  \hspace{0.25cm}= |\bra{R_{nov}\ket{d_1}}^2 \\ \nonumber
&P(R_{nov1} = -1) = |\bra{\widetilde{R_{nov}}\ket{d_1}}^2 
\end{align}
and similarly for document $d_2$.

\subsection{CHSH Inequality for documents using the Trace Method} \label{chsh-trace}
Another way to define the CHSH inequality for documents is by directly calculating the expectation values using the trace rule. According to this rule, expectation value of an observable A in a state $\ket{d}$ is given by
\begin{equation}
\langle A \rangle = tr(A\rho)
\end{equation}
where the quantity $\rho = \ket{d}\bra{d}$ is the density matrix for the state $\ket{d}$.

Let the two documents be represented in the standard basis as follows:

\begin{align}
\ket{D_1} &= a_1\ket{H}_1 + b_1\ket{\widetilde{H}}_1 \\ \nonumber
\ket{D_2} &= a_2\ket{H}_2 + b_2\ket{\widetilde{H}}_2
\end{align}

\hspace{2cm} where $\ket{H}_{1,2} = \begin{pmatrix} 1 \\ 0 \end{pmatrix}$ and $\ket{\widetilde{H}}_{1,2} = \begin{pmatrix} 0 \\ 1 \end{pmatrix}$
Hence, the state vector and the density matrix for a document $\ket{D}$ can be written as:
\begin{align}
\ket{D} = \begin{pmatrix} a \\ b \end{pmatrix} \hspace{1cm} 
\ket{D}\bra{D} = \begin{pmatrix} a_1^2 & a_1b_1 \\ a_1b_1 & b_1^2 \end{pmatrix}
\end{align}
The document representations in another basis are as follows:
\begin{align}
\ket{D_1} &= c_1\ket{N}_1 + d_1\ket{\widetilde{N}}_1 \\ \nonumber
\ket{D_2} &= c_2\ket{N}_2 + d_2\ket{\widetilde{N}}_2
\end{align}
$H$ and $N$ are basically relevance with respect to two relevance dimensions, say Habit and Novelty. We can write the $N$ basis in terms of the $H$ basis (see appendix A) as:

\begin{align}
\ket{N}_1 &= (a_1c_1 + b_1d_1)\ket{H}_1 + (b_1c_1 - a_1d_1)\ket{\widetilde{H}}_1 \\ \nonumber
\ket{\widetilde{N}}_1 &= (a_1d_1 - b_1c_1)\ket{H}_1 + (a_1c_1 + b_1d_1)\ket{\widetilde{H}}_1 \\ \nonumber
\end{align}
and similarly for the second document. 

Thus we get the vector representations for basis states $\ket{N}_1$ and $\ket{\widetilde{N}}_1$ as:

\begin{equation}
\ket{N}_1 = \begin{pmatrix} a_1c_1 + b_1d_1 \\ b_1c_1 - a_1d_1 \end{pmatrix} \hspace{0.5cm}
\ket{\widetilde{N}}_1 = \begin{pmatrix}  a_1d_1 - b_1c_1 \\ a_1c_1 + b_1d_1 \end{pmatrix}
\end{equation}
Now the observables $\boldsymbol{H}$ and $\boldsymbol{N}$ are defined as:

\begin{align}
\boldsymbol{H} &= \ket{H}\bra{H} - \ket{\widetilde{H}}\bra{\widetilde{H}} \\ \nonumber
\boldsymbol{N} &= \ket{H}\bra{N} - \ket{\widetilde{N}}\bra{\widetilde{N}}
\end{align}
where $\ket{H}\bra{H}$ and $\ket{\widetilde{H}}\bra{\widetilde{H}}$ are the projection operators for standard basis vectors with eigen values $1$ and $-1$ respectively. This is the spectral decomposition of the observables. We get $\boldsymbol{H} = \begin{pmatrix} 1 & 0 \\ 0 & -1 \end{pmatrix}$. The matrix for observable $\boldsymbol{N}$ is obtained in terms of the amplitudes $a, b, c$ and $d$. Now the CHSH inequality for the observables $\boldsymbol{H}$ and $\boldsymbol{N}$ acting on the two documents can be written as:

\begin{equation} \label{chsh_expectation}
|\langle \boldsymbol{H}_1\boldsymbol{H}_2 \rangle +
\langle \boldsymbol{H}_1\boldsymbol{N}_2\rangle +
\langle \boldsymbol{N}_1\boldsymbol{H}_2\rangle -
\langle \boldsymbol{N}_1\boldsymbol{N}_2\rangle| \leq 2 
\end{equation}
Here $ \boldsymbol{H}_1\boldsymbol{H}_2$ denotes that we measure the observable $\boldsymbol{H}$ on both the documents. 
In the language of tensor products,

\begin{equation}
\boldsymbol{H}_1 \otimes \boldsymbol{N}_2 \ket{D_1} \otimes \ket{D_2} = \boldsymbol{H}_1 \ket{D_1} \otimes \boldsymbol{N}_2 \ket{D_2}
\end{equation}
And,
\begin{align}
\langle \boldsymbol{H}_1\boldsymbol{N}_2\rangle 
 &= 
 \langle D_1 \otimes D_2 |\boldsymbol{H}_1 \otimes \boldsymbol{N}_2 | D_1 \otimes D_2 \rangle \\ \nonumber
 &= \bra{D_1}\boldsymbol{H}_1\ket{D_1} \bra{D_2}\boldsymbol{N}_2 \ket{D_2} \\ \nonumber
 &= tr(\boldsymbol{H}_1\ket{D_1}\bra{D_1}) \times tr(\boldsymbol{N}_2\ket{D_2}\bra{D_2})
\end{align}
In this way we can directly calculate the expectation values in equation (\ref{chsh_expectation}). As a sample calculation, $tr(\boldsymbol{H}_1\ket{D_1}\bra{D_1}) = tr \bigg(\begin{pmatrix}  1 & 0 \\ 0 & -1 \end{pmatrix} \begin{pmatrix}  a_1^2 & a_1b_1 \\ a_1b_1 & b_1^2 \end{pmatrix} \bigg) = a_1^2 - b_1^2$, where $a_1^2$ and $b_1^2$ are the probabilities of relevance and non-relevance respectively in the standard (Habit) basis.

\subsection{N-Settings Bell Inequality} \label{nsetting-bell}
The CHSH inequality refers to two two-dimensional systems where each system has two measurement settings (or measurement basis). However this can be generalized for systems with multiple settings or basis~\cite{Gisin1999}

\begin{equation} \label{nsetting-inequality}
\sum_{j=1}^{n} \bigg(\sum_{k=1}^{n+1-j} E(A_jB_k) \hspace{0.5cm} - \sum_{k=n+2-j}^{n}E(AjB_k)\bigg) \leq \bigg[\frac{n^2+1}{2} \bigg]
\end{equation}
where $[x]$ denotes the largest integer smaller or equal to $x$. 

For seven relevance dimensions, $n=7$ and the bound is $25$. We can convert equation (\ref{nsetting-inequality}) into its probability form as done in section \ref{chsh-normal}, or use the trace rule to directly calculate the expectation values as done in section \ref{chsh-trace}
\section{Experiment and Results}
Having obtained an equivalent representation of Bell inequalities in section 3, we proceed to substitute the values in the inequalities and test for violation using relevance scores as calculated in section 2. For each query, a user judges several documents to be relevant or non-relevant according to his or her information need. We investigate the correlations between these documents, with each document having multiple decision perspectives, using the Bell Inequalities. We consider the following  types of document pairs to test for Quantum Correlations:
\\

	I) We consider those queries where only two documents are SAT clicked (Satisfied Click - Those documents which are clicked and browsed for at least 30 seconds). Out of $55617$ queries in our dataset, $1702$ queries had exactly two SAT clicked documents. We consider a composite system of these two documents and measure (judge the relevance) along different basis (relevance dimensions) corresponding to each of the Bell inequalities described in sections 3.1, 3.2 and 3.3.
\\

	II) We consider those queries for which we have at least one SAT clicked document. Out of $55617$ queries in our dataset, we find $52936$ queries with at least one SAT clicked document. We then consider a composite system of this SAT clicked document with all the unclicked documents for the query (one by one) and measure (judge the relevance) along different basis (relevance dimensions) corresponding to each of the Bell inequalities described in sections 3.1, 3.2, 3.3.
\\

In both cases, we do not find the violation of the Bell inequalities for any query. While case (I) corresponds to correlated documents and case (II) corresponds to anti-correlated documents, it is to be noted that we are taking a composite system by taking a tensor product of two document states. This, in turn is separable back into the two document states. The reason why Quantum Mechanics violates Bell Inequalities is due to the existence of non-separable states like the Bell States. To get something similar to an entangled state, we consider another type of document pairs:
\\

III) Consider a pair of documents which are listed together for many queries, but are always judged in a correlated manner. That is, if one document of the pair is SAT clicked, the other one is also SAT clicked for that query. And similarly both might be unclicked for another query in which they appear together. Also, we find those documents which are SAT clicked together in half of the queries they occur in, and unclicked in the other half. This corresponds to the following Bell State:
\begin{equation} \label{bell-type-state}
\ket{\psi} = \frac{1}{\sqrt[]{2}}(\ket{RR} + \ket{\widetilde{R}\widetilde{R}})
\end{equation}
We take such pairs of documents to test the Bell inequalities on them. Out of $774$ pairs of documents, no pair show the violation of the inequalities discussed above. 

The composite state of the two documents described in equation(\ref{bell-type-state}) appears to be like an entangled state of the documents - knowing that one document is SAT clicked or not can tell us about the other document. However, one fundamental property of the Bell states is their rotational invariance. Representing a Bell State in any basis, one gets the same probabilities of the two possible outcomes. For example, 
    
\begin{align}
\ket{\psi} &= \frac{1}{\sqrt[]{2}}\big(\ket{HH} + \ket{\widetilde{H}\widetilde{H}}\big) \\ \nonumber
		   &= \frac{1}{\sqrt[]{2}}\big(\ket{TT} + \ket{\widetilde{T}\widetilde{T}}\big)
\end{align}
where H, N and T are relevance with respect to the Habit, Topicality and Novelty basis. One can always hypothetically construct document Hilbert spaces in such a manner that the composite state is rotationally invariant, but that is not the case in the query log data, which is the target of our investigation. 

As a formal test of non-separable states, we perform Schmidt decomposition~\cite{Nielsen} of the composite system of document pairs. We do not find any evidence of non-separable states for any type of document pairs, as described in cases (I), (II) and (III).

\section{Conclusion and Future Work}
We tested Bell inequalities for violation using data from Bing Query logs. Despite the presence of incompatible measurements, Bell inequalities are not violated. However, the incompatibility in measurement applies to the user's cognitive state with respect to a single document. Hence there might exist a joint probability distribution governing user's cognitive state for a pair of documents. The experiments in which the violation of Bell inequality has been reported for cognitive systems, the users are asked to report their judgments on composite states. Hence the joint probabilities can be directly estimated from the judgments. This may result in a ``Conjunction Fallacy'' \cite{Tversky1983-conjunction-fllacy}  due to incompatible decision perspectives, thus violating the monotonicity law of probability by overestimating the joint probability, and therefore violating the Bell inequality. In our dataset, we don't have judgments over the document pairs. That is, the user does not judge a pair of document to be relevant with respect to some dimensions. Instead we have got the probabilities of relevance of a single document with respect to different dimensions. When we use the relevance probability of individual documents to compute the joint probabilities for a pair of documents, we are forced to assume the existence of a joint probability distribution. Thus there might be a possibility of Bell inequality violation if we can obtain data for a pair of documents. For example, users can be asked to rate a document to be relevant with respect to Novelty and another document relevant with respect to Topicality. This would correspond to the $E(R_{nov}, R_{top})$ term in the CHSH inequality. In this case, user's judgment of a document may affect judgment of the other document in the pair. 

Another test of the quantum nature of relevance judgments can be to test the non-contextual inequalities like the KCBS inequality~\cite{kcbs}. Bell inequalities are designed for a composite system with the assumption of locality and realism. The non-contextual inequalities are designed for a single system with multiple measurement perspectives, some of which are incompatible with each other. However, contextuality only exhibits in systems of more than two dimensions. Hence we need to modify our two-dimensional (two decision outcomes - relevant or not relevant) approach to test inequalities like the KCBS inequality. One can also test for violation of the Contextuality-by-Default inequality \cite{Cervantes_Snow_Queen2018,true_contextuality}. This forms part of our future work.  
\appendix
\section{Appendix}
Consider a state vector in two different basis of a two dimensional Hilbert space, $\ket{\psi} = a\ket{A}+b\ket{B} = c\ket{C}+d\ket{D}$
We want to represent the vectors of one basis in terms of the other. To do that, consider the vector orthogonal to $\ket{\psi}$, which is $\ket{\widetilde{\psi}} = b\ket{A}-a\ket{B} = d\ket{C}-c\ket{D}$

Using the above representations, we get 
\begin{equation}\label{c,d}
\ket {C} = c\ket{\psi} + d\ket{\widetilde{\psi}} and 
\ket {D} = d\ket{\psi} - c\ket{\widetilde{\psi}}
\end{equation}

Substituting $\ket{\psi} = a\ket{A}+b\ket{B}$ and $\ket{\widetilde{\psi}} = b\ket{A}-a\ket{B}$ in \ref{c,d}, we get:
\begin{align}
\ket {C} &= (ac+bd)\ket{A} + (bc-ad)\ket{B} \nonumber \\
\ket {D} &= (ad-bc)\ket{A} + (ac+bd)\ket{B}
\end{align}
\section*{Acknowledgment}
This work is funded by the European Union's Horizon 2020 research and innovation programme under the Marie Sklodowska-Curie grant agreement No 721321. We would like to thank Jingfei Li for his help in providing the processed dataset.

\bibliographystyle{splncs04}
\bibliography{bibliography}
\end{document}